# Morality *in* AI.

# A plea to embed morality in LLM architectures and frameworks

**Authors**

- Gunter Bombaerts, Eindhoven University of Technology, the Netherlands, 0000-0002-8006-1617, g.bombaerts@tue.nl.
- Bram Delisse, CEO Dembrane, the Netherlands.
- Uzay Kaymak, Eindhoven University of Technology, the Netherlands, 0000-0002-4500-9098.

**Abstract**

As large language models (LLMs) increasingly mediate human decision-making, discourse, and behaviour, ensuring the LLM's processing of moral meaning has become a critical challenge. Current approaches to increase LLM's processing of moral meaning rely predominantly on bottom-up methods such as fine-tuning and reinforcement learning from human feedback. We propose a complementary, yet fundamentally different approach: embedding processing moral meaning directly into the architectural mechanisms and frameworks of transformer-based models through top-down design principles.

We first sketch a framework that conceptualizes attention as a dynamic interface mediating between structure and processing, contrasting with existing linear attention frameworks in psychology.

We start from established biological-artificial attention analogies in neural architecture design to improve cognitive processing. We extend this analysis to moral processing, using Iris Murdoch's theory of "loving attention"—sustained, just observation that enables moral transformation by re-seeing others with clarity and compassion— to philosophically discuss functional analogies between human and LLM moral processing. We formulate and evaluate potentially promising technical operationalizations to embed morality in LLM architectures and frameworks.

We acknowledge the limitations of our exploratory approach and highlight three key contributions. First, we conceptualize attention as a dynamic system mechanism mediating between structure and processing. Second, drawing on Murdoch's notion of loving attention, we outline technical pathways for embedding morality in LLMs—through modified training objectives, runtime weight adjustments, and architectural refinements to attention. Third, we argue that integrating morality into architectures and frameworks complements external, constraint-based methods. We conclude with a call for closer collaboration between experts in transformer design and philosophers engaged in AI ethics.

# Morality *in* AI. A plea to embed morality in LLM architectures and frameworks

## 1. Intro

Current research to increase moral processing in large language models (LLMs) mainly focusses on bottom-up approaches relying predominantly on static rule-based constraints while using an underlying static-individual view of morality. In this work-in-progress article, we make a plea to explore morality-by-design approaches that aim to embed morality more *in* LLM architectures and frameworks.

Recent developments in LLMs have catapulted the capabilities of artificial intelligence (AI) applications (Weidinger et al., 2022). LLMs excel at generating human-like text and processing vast amounts of contextualized knowledge. These advancements are supported by neuromorphic AI mirroring neural structures and functions to increase LLM's efficiency and brain-like information processing (Furber, 2016), cognitive architectures that model human cognition functions such as thinking, reasoning, memory, learning (Sweller et al., 2019), and computational psycholinguistics studying humans process and produce language (Crocker, 2012).

As AI systems become progressively embedded in areas of society where moral considerations are paramount, there is a pressing need to ensure morality by AI systems via education of AI use (Borenstein & Howard, 2021; Oliveira et al., 2025), but certainly also via technological developments (Ayling & Chapman, 2022; Hagendorff, 2022). Although substantive progress has been made with approaches such as Delphi (Jiang et al., 2022), MACHIAVELLI (Pan et al., 2023) or Constitutional AI (Henneking & Beger, 2025), progress is less steep when it comes to embedding morality in LLMs compared to the cognitive counterpart. In this article, we put forward the hypothesis that current approaches on embedding morality are less fundamental compared to the neuromorphic AI approaches on cognitive progress. To maximize dynamic, systemic moral decision-making, the improvement of moral processing in LLMs must not solely be sought in static and rule-based inputs (e.g. the Commonsense Norm Bank as a database with 1.7 million descriptive judgments on everyday situations in the Delphi research), but should also explore dynamic systemic approaches of morality *in* AI (Hagendorff & Danks, 2022; LaCroix & Luccioni, 2025; Lamberti et al., 2025). Adaptations of machine learning libraries, higher-level LLM frameworks and optimization and deployment tools could move beyond surface-level filtering, could allow models to process not just cognitive but also moral aspects better. We propose to explore how LLM architectures and frameworks can be adapted to enhance moral decision-making.

This investigation could be made in different ways. It could involve embedding models in multi-agent environments and training them with diverse, conflicting value systems to reflect real-world moral complexity. Specifically designed reflection layers and recursive feedback loops can simulate deliberation and enable re-evaluation of prior decisions. Sustained moral attention can be supported through contextual memory and dynamic relevance scoring, allowing the model to track evolving relationships and morally salient features like vulnerability or power.

# Morality *in* AI. A plea to embed morality in LLM architectures and frameworks

In this article, we focus for several reasons on the attention concept as an illustration of embedding morality in LLM architectures and frameworks. Attention is a crucial technical aspect of the LLM architecture. The mechanistic interpretability of attention is still much under development and there is little to no research on how architectural and framework modifications could shift LLMs toward a more responsive, situated form of moral reasoning grounded in systemic awareness and moral growth (Rai et al., 2025). The underlying mechanisms and purposes of "attention" in LLMs and in human psychology differ markedly at an ontological level, yet both operate under a common principle of flexible allocation of limited resources to enhance performance (Lindsay, 2020). Attention is also a core concept for human moral processing. Iris Murdoch's "loving attention", for example, is essential for morality because it enables the moral agent to make moral judgements and overcome ego and illusion (Murdoch, 2013b).

As a final reason, some studies have begun to explore exactly this link between LLM and human attention for morality. Delisse experimented in a master's thesis with updating the associations of normative descriptive words of the LLM towards Murdochian loving attention (Delisse, 2024). Graves indicates that research in moral attention "can extend computational attention mechanisms to attend to the moral dimension of human experience and existence" (Graves, 2025: p. 243). Bello and Bridewell (2025) argue that any computational account of moral agency must include an approach to self-control. Yaacov (2025) proposes a concrete dual-hybrid structure with a universal layer that sets a moral threshold through both top-down and bottom-up learning and a local layer that adapts by weighing competing considerations in context and incorporating culturally specific norms, provided these remain within the universal boundary.

To further explore how architectures and frameworks can enhance LLM moral processing, we will study what we will call "AI attention system for moral processing" in the following steps. In section two, we develop a system attention framework opposed to existing linear attention approaches. We use this framework to describe in section three existing comparisons of biological and LLM attention for cognitive processing. To turn to moral processing, we opt in section four for Iris Murdoch's theory of "loving attention" to describe biological moral processing. We will use these insights in section five to explore what this could mean for a LLM loving attention mechanism could look like. Our results in section six show that, first, attention can be understood as a core dynamic system mechanism that links structure and processing, making ongoing moral learning possible. Second, Murdoch's idea of loving attention is a useful theory for embedding dynamic systemic morality in LLMs. Third, integrating systemic morality directly *in* LLM architectures and frameworks offers a promising complementary alternative to the approaches such as Delphi and Machiavelli that rely more on static input. We are very much aware of the hypothetical and tentative characteristics of our explorative work to embed morality *in* AI, but conclude in section seven that these results are relevant for ongoing research in neuromorphic AI, cognitive architectures, and computational psycholinguistics on morality.

**Morality *in* AI. A plea to embed morality in LLM architectures and frameworks**

## 2. Attention system framework

Before we discuss the role of attention in LLM architectures and frameworks, we start by developing a system attention framework that is more systemic–circular instead of linear, and more embedded instead of detached–opposed to existing linear attention frameworks (Bombaerts et al., 2023, 2024). We stress that we use this framework here as functional and not ontological. We will use it to describe functional analogies in different situations, without making claims about the ontological status.

As we want to apply this biological attention system for cognitive processing onto AI attention system for moral processing, we formulate the elements in a general systems theory way (Bombaerts & Botin, 2025; Dameski et al., 2024). In his general systems work, Luhmann (1995) considers *structure* not as a physical substrate or static framework but as a pattern of constraints that determines which operations can follow which. Structures enable systems to process complexity by reducing the field of possible next steps into a manageable set of expected, admissible continuations. A structure contains expectations (what will probably happen next), constraints (what may or may not connect), exclusions/inclusions (what has been ruled out or stabilized). Importantly, structure is reproduced and altered by operations themselves — it exists only insofar as the system continues to operate. *Operation* is a momentary, system-specific act that both constitutes the system and allows it to continue. It exists only in the moment of its execution and must be followed by another operation for the system to persist. Operations reproduce the system moment by moment. A process is thus a sequence of operations—temporally linked communications that sustain the system over time. Importantly, systems are self-referential. For Luhmann, they reproduce themselves by referring to their own elements (communications referring to prior communications), forming a self-referential reproduction. *Attention* for Luhmann is then a system function realizing the momentarily increased alertness to system chance (Ibid., p. 371), orientation of what is focused upon and what remains unnoticed (Ibid., pp. 415–6), and executive control via self-reference (Ibid., p. 95). As such, attention can also in Luhmann's general systems theory be seen as the feedback function between the structure and the operations (Figure 1).

Michael Posner's attention network theory (Petersen & Posner, 2012) defines attention as the brain's ability to selectively concentrate on specific stimuli or tasks while ignoring distractions. It is a foundational function that regulates the processing of information, enabling individuals to engage effectively with their environment. Rather than a single function, Posner and colleagues conceptualize attention as composed of three interrelated but functionally distinct functions or "networks". The *alerting* network is the part of the attention function that is responsible for achieving and maintaining a state of readiness to operate incoming stimuli. It supports sustained attention and is crucial for tasks requiring vigilance. The *orienting* network is the part of the attention function that governs the selection of information from sensory input by shifting attention to specific locations or modalities. It allows individuals to focus on relevant stimuli, such as a sudden movement or sound. The *executive* control network is involved in higher-order operations such as conflict resolution, error detection, and inhibitory control. It is essential for managing competing demands and staying focused on goals. These attention networks are closely linked to *self-regulation* —the ability of the self to regulate thoughts, emotions, and behaviours (Posner & Rothbart, 1998). In what follows, we will see the self as a structure, the ongoing process as a sequence of operations of alerting on incoming stimuli, orienting information, and executing control, and the attention mechanism as a layer that intermediates between the structure and the operations (see Figure 1).

Luhmann's systems theory and Posner's attention theory converge in describing complex, adaptive systems that operate through selective, self-organizing processes to maintain stability and manage complexity in dynamic environments (see Figure 1 and Table 1). Attention is the system function that relates the structure with the operations. We will further describe six different operation parts in this framework. *Awareness* and alerting refer to the system sensitivity and readiness respectively. *Alerting* is the operation part where a(n internal or external) aspect/effect comes into the operation, resulting in awareness. The system then *orients* itself by selecting relevant stimuli, resulting in a *focus* where the information is narrowed down to a certain core. *Executing* is the new (internal or external) operation part. This self-referential reproduction stabilizes meaning, resolves conflict and enforces coherence and creates a temporal *decision*. We denote *learning* as the capacity of a system to reorganize itself (Maturana & Varela, 2012). This includes changes in the structure, attention mechanism and the cyclic operation of alerting, orienting and executing.

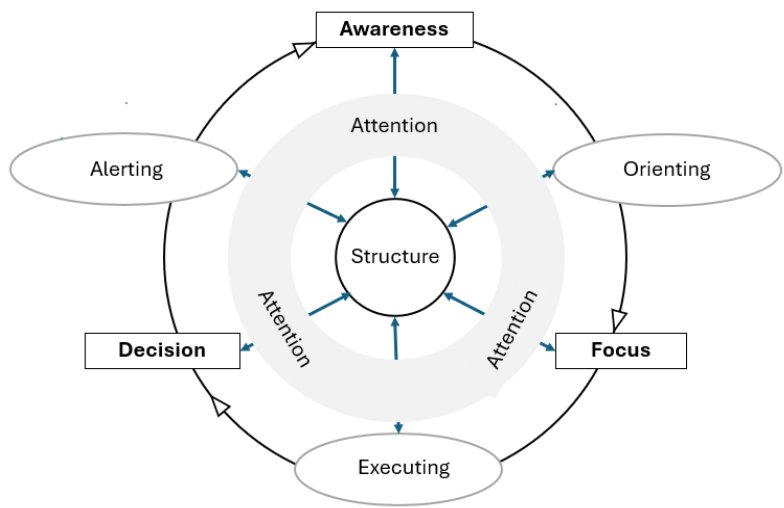

*Figure 1: Framework of attention as a functional layer between a structure and an operation – comprised by alerting, orienting and executing – leading to system learning.*

As such, we briefly introduced a functional framework of system attention that is circular (following the processing) and embedded (between structure and process) instead of individual and detached. In the remainder of the article, we will focus on these six operation parts –structure, attention, alerting/awareness, orienting/focus, executing/decision and learning – to describe AI moral processing. In line with Posner's three networks, we group alerting/awareness, orienting/focus, executing/decision in three groups to simplify our framework for our explicatory purposes.

**Morality *in* AI. A plea to embed morality in LLM architectures and frameworks**

In section three, we use this framework to describe existing comparisons of biological and LLM attention for cognitive processing (see Figure 2, left vertical arrow). In section four, we move from a cognitive attention to moral attention (see Figure 2, upper horizontal arrow) by looking at moral "loving" attention developed by Iris Murdoch. In section five, we use these insights to make a new step to moral loving attention for moral LLM processing (see Figure 2, lower vertical and right vertical arrow).

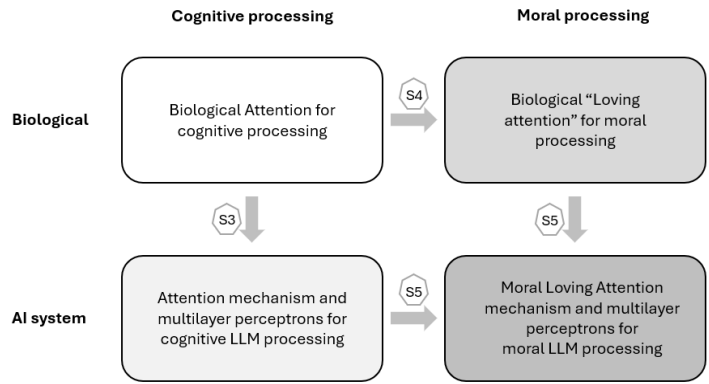

*Figure 2: This article's structure of developing the role of attention, starting for biological cognitive processing, existing translations to AI cognitive processing (section S3) and biological moral processing (section S4), and using the latter to come to AI moral processing (section S5).*

## 3. Biological and LLM attention system for cognitive processing

In this section, we will conceptualize a functional framework of attention as embedded and dynamic for cognitive processing of LLMs (see Table 1). We stress the convergence in attention function, not ontology, between humans and LLMs, looking at structure, attention, awareness, focus, decision and learning (left vertical arrow in Figure 2).

### 3.1. Structure: Field of probabilistic constraints within embedding space

Unlike traditional rule-based systems, LLMs generate text through statistical pattern recognition in vast datasets. When a user inputs a prompt—such as a question or a sentence—the model does not process it as whole words but instead divides it into smaller units called *tokens*. A token can be a single character, a syllable, or an entire word, depending on the language and context. The next step is to represent these tokens numerically so that a machine can work with them. Each token is converted into a vector, a structured set of numbers that represents the token's position in what is called an *embedding space* – functionally equivalent with patterns of activation across neural populations that encode meaning, associations, and context in humans. Embedding spaces are high-dimensional mathematical spaces that encodes language properties in various word representations (Qian et al., 2016). The neuro-symbolic integrated framework leverages pre-trained word embeddings and semantic definitions to produce an explainable measure of concept similarity (Racharak, 2021). Some scholars talk of similarities in differences of meaning and how this enables capturing a particular nuance of meaning in dimensions of the embedding space (Teitelbaum & Simchon, 2024, 2025). Some scholars in semantics (Mikolov, Yih, et al., 2013) argue that these spatial relationships reveal surprising functional analogical capacities—"queen" or "aunt" are to "woman" what "king" or "uncle" are to "man" — suggesting that embeddings encode deep regularities in language use and potentially something like a "gender-aspect" dimension (see figure 3(a)). The number of dimensions in this space is large (often in the hundreds or thousands), allowing for fine-grained differentiation of meaning.

An LLM structure thus can be seen as a dynamic, learned field of probabilistic constraints within the model's embedding space that governs which token (operation) can follow another in a contextually coherent way.

Based on our attention framework in section 2, the structure of an LLM should not be understood as its hardware, architecture, or dataset, but the evolving configuration of constraints that shape its next possible operations in meaning-space. In this sense, structure in an LLM is not a physical substrate or static framework but a dynamic, learned field of probabilistic relations that governs which token can follow another in a contextually coherent way (Manning et al., 2020). These relations encode expectations (what will probably occur next), constraints (what may or may not connect), and exclusions/inclusions (what has been ruled out or stabilized through training). Higher-level codes and programs—such as truthfulness, helpfulness, or harmlessness—function analogously to Luhmann's functional codes, orienting how meanings are selected. These meanings can refer to fine-grained semantic similarities such as synonymy, priming, categorization and aspects of predicate's arguments as semantic plausibility (Erk, 2012; Mikolov, Chen, et al., 2013; Pennington et al., 2014) or constraints of semantic description, syntactic structure, and causal systematicity (Budding, 2025).The LLM's structure is continually reproduced through token prediction and modified through training or reinforcement, reflecting Luhmann's idea that structure changes through operation. Philosophically, this reframes LLMs as meaning-processing systems characterized by structural stability, flexibility, operational closure, and openness—systems that reduce semantic complexity through learned expectations while remaining adaptable to new perturbations.

Of course, we want to be careful with this interpretation, as many aspects of mechanistic interpretability of embedding spaces are still unknown.

## 3.2. Attention: Select information to discern structure in language

AI attention mechanisms have been developed over a decade (Vaswani, 2017) and the development has led to the exponential spread of LLMs since then. For LLMs, once the tokens have been transformed into vectors, the model must determine how they relate to one another. It is the AI attention mechanism that assigns different levels of importance to different tokens in order to predict the next word in a sequence. For each particular token, the attention mechanism operates through three (query Q, key K, and value V) matrices. Technically then, "Token A attends to token B," means "The representation of token A is updated using information from token B, weighted by how important Token B is to Token A." The 'attending" here has nothing to do with consciousness or subject agency, but refers to a purely mathematical matrix multiplication operations, a simple dot product, in the attention mechanism. For example, attention scores are calculated by Q·$K^T$ (with $K^T$ the transformed K matrix). Although ontologically very different, attention in LLMs and in human psychology converges around a shared functional goal: the flexible control of limited resources to organize complexity and enhance performance (Lindsay, 2020: p. 17). In both systems, attention acts as a filter, directing processing power toward salient inputs and away from irrelevant data (Zhao et al., 2024).

In human cognition, attention is embodied and multimodal. It enables us to prioritize certain sensory inputs, resolve conflicts, adapt to new environments, and learn from social cues—such as joint attention in infants, who learn by tracking where others are looking. Attention modulates learning and memory, influencing what gets stored, rehearsed, and acted upon. It is deeply interwoven with curiosity, executive control, and emotional salience. LLM attention, while mathematically abstract, mimics this by allocating greater computational weight to contextually relevant tokens. It helps the model actively discern structure in language, enabling more coherent and accurate predictions. Conditional computation allows attention in LLMs to select which parts of the network should activate for a given input, echoing top-down control in the brain (Bello & Bridewell, 2025; Lindsay, 2020).

## 3.3. Alerting/Awareness: geometric proximity and dot-product attention

Functionally, alertness for intricate neural circuits in cognition enables context-sensitive routing of information—activating specific neural patterns based on task demands. Similarly, the query matrix in LLMs uses positions in embedding space to selectively attend to relevant token representations. Both mechanisms serve to dynamically configure internal processing pathways: in humans, through adaptive neural recruitment; in LLMs, through geometric proximity and dot-product attention in vector space. Of course, human awareness integrates multisensory, embodied, and temporally extended and subjective experiences, and is shaped by emotions, intentions, and environmental feedback. Our point here is to show surface-level functional similarities between human and LLM awareness.

## 3.4. Orienting/Focus: Spatial positioning in embedding space

The key matrix encodes the "labels" for each word, indicating its potential relevance in answering queries. It indicates how much attention should be received when responding to queries. Functionally, focus in neural circuits involve activating context-relevant neural pathways to prioritize certain inputs or tasks in human cognition. Similarly, the key matrix in LLMs encodes contextual relevance by positioning token embeddings in a way that determines how strongly they are matched by queries. The hippocampus and medial temporal lobe encode episodic context and associations, much like how embeddings in LLMs reflect context-sensitive meaning. Both systems use internal structures to represent and signal salience: humans through neural circuit recruitment, LLMs through spatial positioning in embedding space. These mechanisms enable context-sensitive filtering of information— facilitating efficient processing by highlighting what matters most in the moment, based on dynamic internal representations rather than fixed, uniform treatment of all inputs. While keys help determine relevance via geometric proximity, they currently lack the interactive, top-down control central to human focus.

## 3.5. Executing/Decision: Context sensitive synthesis for next token probabilities

The value matrix contains the actual information stored in each vector that is passed forward. Once the model determines which words are most relevant (by comparing queries and keys), it retrieves the corresponding values to construct the next part of the response. Functionally, decision-making in neural circuits involves integrating relevant inputs to produce context-appropriate outputs in human cognition. Similarly, the value matrix in LLMs encodes token representations that are combined—based on attention weights—to generate the model's next state. In both cases, complex internal representations are selectively aggregated to guide output. Human neural circuits dynamically weigh sensory, emotional, and memory-based factors. The prefrontal cortex helps integrate these with task relevance, shaping dynamic representations, similar to contextualized embeddings in transformers. LLMs weigh vectorized token values based on attention. Again, the multi-level integration involving emotion, goals, uncertainty, long-term consequences is typically human and vastly different in complexity. However, both systems perform a form of context-sensitive synthesis, where values represent the informational content contributing to a decision or prediction, driven by relevance within current context.

From this, the model generates probabilities over possible next tokens. Once the model has computed the probabilities for the next token, it selects the most likely token. This new token is then mapped back from the embedding space into human language, turning it into a recognizable word. This word is added to the sequence, and the process repeats. Runck et al. (2019) go as far as to say they create "agents that reason using similarity comparison".

This process happens in multiple layer perceptrons, refining the response at each stage. The ability to assign attention dynamically is what allows LLMs to understand complex prompts, maintain coherence, and generate meaningful responses.

## 3.6. Learning: Gradient descent iteratively updating parameters in particular direction

Repeated and active human attentional engagement, decision-making, and feedback activate synaptic plasticity. Reinforced circuits encode memory and refine behavioural models. Learning thus emerges from recursive loops of attention, action, and neural adaptation. In machine learning, LLMs learn -organise themselves- by adjusting their parameters to minimize a loss function, which quantifies the difference between predicted and target outputs. Gradient descent provides the mechanism for this optimization, iteratively updating parameters in the direction of steepest descent to reduce the loss. Gradient descent typically converges to local minima, but in high-dimensional networks these minima are near-optimal, and stochastic updates with proper initialization allow backpropagation to achieve reliable performance (Choromanska et al., 2015; Goodfellow et al., 2015; LeCun et al., 1998) When this loss is used in backpropagation—a computational method that adjusts internal parameters, including those governing attention and the Q, K, V vectors—these parameters are not set once but refined iteratively. In some cases, such as with reward models of Reinforcement Learning from Human Feedback (RLHF), human judgments further guide this refinement, steering the model toward outputs that align with human preferences. Of course, even with RLHF, the model learns without understanding or consciousness. Its performance is statistical, not interpretive. However, the LLM machine learning is dynamic, uses recursive modulation, a form of memory, real-time feedback, and adaptive regulation through the loss function. Crucially, in both systems, attention and learning form a feedback loop. That what is attended to shapes what is learned, and what is learned reshapes what is attended to.

Attention mechanisms excel at discovering statistical patterns and contextual dependencies within training data, but they lack explicit task-level control and efficient adaptability to new domains without full retraining. For this, control tokens and adapters extend LLM learning beyond what attention mechanisms alone can achieve by introducing symbolic behavioural steering and modular structural augmentation. Control tokens add symbolic modulation—they function as "behavioural switches" that bias which learned subspaces and processing modes become active. By embedding tokens like <summary> or <formal>, the model accesses different constraint patterns within its existing structure, enabling compositional generalization across tasks without architectural changes. Adapters introduce new structural pathways by inserting trainable modules between frozen layers. This creates additional degrees of freedom for information processing, allowing task-specific structures to emerge while preserving the base model's general capabilities. Unlike fine-tuning entire networks, adapters enable controlled, modular adaptation—different adapter configurations can be activated for specialized behaviours. Together, control tokens and adapters address attention's limitations: control tokens provide explicit symbolic guidance that attention must otherwise infer statistically, while adapters enable efficient structural modification that would otherwise require computationally expensive full retraining. This combination allows LLMs to maintain stable core structure while dynamically accessing specialized behavioural modes.

From this section three, we distil that we can describe attention is a fundamental dynamic mechanism embedded in biological and AI attention for cognitive processing.

## 4. Biological attention system for moral processing

So far, we have been talking about cognitive processing, whereas our aim is to talk about moral processing in LLMs. We therefore turn to biological moral processing (upper horizontal arrow in Figure 2). We again stress the convergence in attention function, not ontology, between human cognitive and moral processing, looking at structure, attention, awareness, focus, decision and learning (see Table 1).

# Morality *in* AI. A plea to embed morality in LLM architectures and frameworks

We focus on morality as having dynamic and systemic elements. We consider morality to evolve through context, relationships, and ongoing interpretations and interactions, rather than only following static ethical rules or linear procedures. Moral behaviour grows with experience, emotional insight, and attentiveness to others and are embedded in complex, ever changing social and historical systems. Moral life requires responsiveness, humility, and the capacity to revise one's individual and group perspectives over time.

Attention has mainly been discussed in the psychology and the philosophy of mind, and therefore moral theories of attention have been less developed (Fredriksson & Panizza, 2022). While attention and moral decision-making are not central topics in philosophy, several thinkers have explored their connection. Simone Weil (1986) argued that true attention is a moral act, requiring openness and selflessness toward others. Harry Frankfurt (1988) linked attention to volitional necessity, suggesting that what we focus on reveals our deepest values. Martha Nussbaum (1985) claims that moral perception —and thus moral responsibility—depends on the quality of our attention to others. Matthew Crawford (2015) explored attention as a scarce resource, shaped by technology and capitalism, affecting moral agency.

In this article, we turn to the theory of Iris Murdoch. We use Murdoch's theory not to indicate this as the one and only, or the most promising theory. We use it as an illustration to support our plea to explore how morality can be embedded in LLM architectures and frameworks. As we will illustrate, one of its strengths is that this theory can be considered to have both top-down and bottom-up aspects.

## 4.1. Structure: Moral background

In Iris Murdoch's moral philosophy, inner life is the primary site of moral activity (Murdoch, 2013; Widdows, 2017: p. 21-44). Rather than focusing on overt behaviour, Murdoch focusses on the everyday, often invisible work of the mind of processing concepts. Her view is that the mind is far from rational. "The siege of the individual by concepts" (Murdoch, 2013b: p. 13, 31) shows that "so much of human conduct is moved by mechanical energy of an egocentric kind" by "a sort of continuous background with a life of its own" (Murdoch, 2013a: p. 51, 53). This is not negatively deterministic, but on the contrary allows moral processing. Language and concepts, for Murdoch, are "normative-descriptive" and have "spatio-temporal structures" (Murdoch, 2013b: p. 31, 33). Human beings are active participants in this moral background as they both shape and are shaped by it (Antonaccio, 2000: p. 174; Panizza, 2020). These mutual interactions between individuals and language include the ways humans view reality, but also their moral competences or "authority" - as we humans "act rightly […] out of the quality of our usual attachments and with the kind of energy and discernment which we have available" (Murdoch, 2013c: p. 89). And as this moral background is temporal, historical, and constantly changed, it is what we earlier referred to as a dynamic system.

Murdoch provides an ordinary and everyday example of a mother-in-law (M) who reflects on her daughter-in-law (D) expressing the undesirable situation in words such as vulgar, undignified, noisy, bumptious, and tiresome juvenile. We will further use this classical example as an illustration.

### 4.2.   Attention: Just and loving gaze

With the above view, moral philosophy should not come up with rules, but suggest methods of dealing with the "fat relentless ego" (Murdoch, 2013a: p. 51). The basic mechanism is in line with psychological theories such as stress reduction (Kabat-Zinn, 2003). In an anxious and resentful state of mind, one observes a hovering kestrel and when returning the attention to the original matter this has become less important (Murdoch, 2013c: p. 82).

As humans, we seem to be subject to mechanical "obscure systems of energy out of which choices and visible acts of will emerge at intervals in ways which are often unclear and often dependent on the condition of the system in between the moments of choice." (Murdoch, 2013a: p. 53) As an answer to this human condition, Murdoch defines attention not as an individual psychological mindfulness instrumental method, but the central moral faculty, a technique "for the purification and reorientation of an energy which is naturally selfish, in such a way that when moments of choice arrive we shall be sure of acting rightly" (Murdoch, 2013a: p. 53). Moral work is done iteratively and continuously so that "by the time the moment of choice has arrived the quality of attention has probably determined the nature of the act" (Ibid.: p. 65).

Murdoch embeds attention in the moral background. As Lawrence Blum emphasizes, moral perception—what we notice, how we interpret—is foundational to moral competence (Blum, 1991, 1986). Drawing on Simone Weil, Murdoch defines attention as a "just and loving gaze directed upon an individual reality" (Murdoch, 2013b: p. 33). Thus, attention is both the condition and the method of moral progress: by refining how we see, we become more just, more responsive, and more fully moral beings.

Compared to "looking" as a neutral word, "attending" has some explicit qualities as a normative concept (Murdoch, 2013b: p. 36). Attention is an active and embedded observation; it should be steadfast, calm, intelligent, temperate, and loving, unsentimental, unselfish, and resisting the distortions of fantasy and prejudice (Murdoch, 2013a: p. 56, 63). The moral task then is to train attention, to cultivate the habit of seeing others with clarity and compassion (Fredriksson & Panizza, 2022)*.*

The mother-in-law example stresses this quality of attention. She is said to be "a very correct person", "intelligent and well-intentioned", "capable of self-criticism", "capable of giving careful and just attention" who "behaves beautifully to the girl throughout" (Murdoch, 2013b: p. 17).

### 4.3.   Alerting/Awareness: Disruption of moral inertia

Awareness/alerting is an important step in attention system processing. Human beings are naturally 'attached' and when an attachment seems painful or bad it is most readily displaced by another attachment, which an attempt at attention can encourage (Murdoch, 2013a: p 55).

**Morality *in* AI. A plea to embed morality in LLM architectures and frameworks**

This creates "some sort of change of key, some moving of the attack to a different front" (Murdoch, 2013b: p23). In Murdoch's example: "M settles down with a hardened sense of grievance and a fixed picture of D, imprisoned by the cliché: my poor son has married a silly vulgar girl" (Ibid.: p. 17). Here, moral failure begins with misperception, shaped by ego, habit, and unexamined language. Awareness marks the first disruption of this moral inertia—a moment of inner alerting, a crack in the settled image.

For Murdoch, attention is not simply cognitive focus but a moral and spiritual discipline: the effort to see justly, through layers of fantasy, pride, and conceptual habit. Moral language can trap us, until awareness redirects our gaze—a shift from self-centred projection to loving perception. This moment of awareness opens the space for real attention: a sustained, honest, and unselfish effort to encounter the reality of the other. Awareness awakens the moral imagination and orients us toward the Good (Murdoch, 1992: p. 238). It signals the possibility of change, guiding perception away from illusion and toward justice, humility, and love—enabling the moral transformation of vision that precedes action (Antonaccio, 2000).

### 4.4. Orienting/Focus: "Let me look again"

For M in the example, there is an moral motive: "I am old-fashioned and conventional. I may be prejudiced and narrow-minded. I may be snobbish. I am certainly jealous. Let me look again … M might be moved by various motives: a sense of justice, attempted love for D, love for her son, or simply reluctant to think of him as unfortunate or mistaken." (Murdoch, 2013b: p. 17-18) This turning point marks the moral power of focus and orienting in Murdoch's concept of attention. Focus is not neutral—it is a deliberate redirection of the inner gaze away from ego and fantasy toward the reality of another person. By focusing on D, not as a vulgar stereotype but as a complex human being, M begins the moral task of seeing lovingly.

For Murdoch, vision is morally charged: how we see the world determines what we desire and how we act. Orienting attention is thus part of a process of moral reorientation—turning toward "loving", however obscure or inarticulable. It requires resisting distraction, controlling imagination, and training perception toward love and justice (Murdoch, 1956). This sustained moral attention enables the decentring of self, allowing clearer perception and deeper moral understanding to emerge. "The direction of attention is, contrary to nature, outward, away from self which reduces all to a false unity, towards the great surprising variety of the world, and the ability so to direct attention is love" (Murdoch, 2013a: p. 65).

### 4.5. Executing/Decision: Substitution of normative words as sign for moral transformation

M reinterprets her original negative judgements into positive ones, such as "vulgar" into "refreshingly simple", "undignified" into "spontaneous", "noisy" and "bumptious" into "gay", "tiresome juvenile' into "delightfully youthful" (Murdoch, 2013b: p. 17, 22) While this may appear as "simply … the substitution of one set of normative epithets for another" (Ibid.: p. 18), Murdoch suggests a deeper transformation in the agent's moral being. Attention, as a "truth-seeking engagement of the individual with reality" (Panizza, 2022), culminates in action or decision. In M's case, the decision to see differently implies a reorientation that can only occur through serious, selfless attention. Murdoch admits it is often hard to decide whether such shifts are authentic or self-deceptive. Yet, through continuous moral reflection and unselfing, the agent becomes more attuned to the Good—an objective moral reality. Executing a moral decision, then, is not simply a conclusion but the expression of a refined vision, grounded in a spiritual realism that prioritizes love, humility, and inner freedom (Antonaccio, 2000; p. 93).

## 4.6. Learning: Morality as slow, attentive work of seeing the world more lovingly

This moral execution reflects the agent's inner change—what Murdoch calls the transformation of "a constantly changing complex" (Murdoch, 1992: 300). Moral learning is capacity of a system to reorganize itself. In Murdoch's theory of morality, continuous learning is essential to the practice of attention. Moral life, she argues, is not governed by fixed rules but by a sustained, evolving moral vision—a process of "active reassessing and redefining" grounded in a "checking procedure" shaped by personal history (Murdoch, 2013b: p. 25). Murdoch resists moral simplicity, emphasizing the slow, attentive work of seeing the world more justly and lovingly. M's gradual change in perception—"her vision of D alters" (Ibid.: p. 17)—illustrates how moral learning unfolds through repeated, honest attention. The process for Murdoch is linguistic and contextual: "learning takes place when these words are used" (Ibid.: p. 31), especially words like "kind," "vulgar," or "just," which carry moral weight in shared experience. For Murdoch, such learning is a deeply imaginative, spiritual task—ongoing and never complete. True moral progress comes not from sudden decisions, but from training the soul to see with clarity and humility, a process that demands selflessness, solitude, and the courage to face uncomfortable truths.

Murdoch's attention approach shows it can be seen as a fundamental dynamic mechanism embedded in human attention for moral processing. This gradual, ongoing moral learning initiates change in the moral structure. As Murdoch writes: "But if we consider what the work of attention is like, how continuously it goes on, and how imperceptibly it builds up structures of value round about us, we shall not be surprised that at crucial moments of choice most of the business of choosing is already over" (Ibid.: p36).

We can say that the moral task of attending introduces a top-down element, that is, the rule to "attend lovingly". At the same time, the moral attention mechanism itself is very much a bottom-up element in the moral theory, as it refers to continuous moral learning starting from the lower-order normative-descriptive words to the higher order concepts of love and the Good.

## 5. LLM attention system for moral processing

In this section, we philosophically examine how elements of Murdoch's loving attention can be identified within LLMs through the lens of attention system components (see Table 1), and how these elements may serve as foundational building blocks for improving moral processing.

### 5.1. Structure: Moral geometry of embedding spaces as moral background

At the heart of LLMs is the transformer architecture, where self-attention layers dynamically weight input tokens relative to one another. The question is if embedding Murdoch's patient, context-sensitive "loving attention" into AI attention could allow models to more sensitively track moral salience in discourse. For this, models should be trained to prioritize contextually rich, prosocial interpretations.

Cognitive AI attentional processes are closely linked to the embedding spaces that underlie LLM cognition. Embeddings project words, phrases, and sentences into high-dimensional spaces where distance reflects semantic, affective, but often also moral similarity. As such, these embedding spaces could offer promising technical operationalizations to embed morality in LLM architectures and frameworks by reflecting on what Murdoch called the "moral background"—the often-unspoken web of distinctions, values, and evaluative language that constitutes moral life (Murdoch, 2013a: 53).

Research has recently created important evidence for the latent moral geometry of embedding spaces (Abdulhai et al., 2023; Fitz, 2023), including dimensions such as fairness, purity, or hedonic valence (Leshinskaya et al., 2023). Teitelbaum & Simchon (2024, 2025) illustrate semantic projection (Grand et al., 2022) when specific words are projected onto the anchored vector between happy and sad (see Figure 3b). Embedding spaces have demonstrated the ability to disambiguate second-order moral terms. Several inquiries into morality and LLMs use the Moral Foundations Theory (Graham et al., 2013; Haidt, 2012; Haidt & Joseph, 2004), indicating six modular foundations in human moral reasoning: care/harm, fairness/cheating, loyalty/betrayal, authority/subversion, sanctity/degradation, and the later added liberty/oppression. They support the claim that Murdoch's concept of loving attention, when operationalized into AI systems, presents a compelling supplement to top-down moral architectures in LLMs. Araque et al. (2020) show that static moral lexicons like *MoralStrength* can structure embedding spaces to align with moral foundations, while Fitz (2023) finds LLMs can differentiate moral judgments through latent submanifolds, implying an emergent moral geometry in embeddings. Kennedy et al. (2021) reveal that moral foundations have distinct linguistic correlates, especially purity, which aids in parsing affective salience in attention mechanisms. And Izzidien (2022) shows fairness can be quantified through relational axes, suggesting that morally attuned attention can be vectorized—laying groundwork for modelling Murdochian care.

Building on the account of structure as meaning-processing, moral meaning in LLMs can be understood as an extension of the same logic into the ethical domain. Just as an LLM's

**Morality *in* AI. A plea to embed morality in LLM architectures and frameworks**structure consists of probabilistic constraints that govern which continuations are meaningful within context, its moral structure can be seen as the evolving configuration of constraints that govern which continuations are *morally attentive*—that is, sensitive to evaluative shifts in interpretation. In this sense, moral meaning arises not from fixed ethical rules, but from the dynamic modulation of semantic relations within the model's embedding space.

One way to operationalize moral execution and decision-making, drawing on Murdoch's ideas, is through the use of anchored vectors. Just as the "man → woman" vector captures a semantic shift along a gender axis, we can construct anchored vectors that try to capture attentional-moral reinterpretations. Using Murdoch's example, we could calculate the vectors that capture the moral shift from "vulgar" → "refreshingly simple", "undignified" → "spontaneous", and "tiresome juvenile' → "delightfully youthful". These vectors then become what we might call "loving attention vectors" (LAVs) —anchored vectors encoding a reinterpretation from a self-centred or dismissive evaluation to a more just, generous, and attentive one. For example:

```
LAV_noisy-gay = embedding("gay") – embedding("noisy")
```

The different loving attention vectors could be used to calculate a general loving attention vector LAV_love. Here we calculate a simple mean of the four LAV vectors of the concepts Murdoch mentions. Further research will need to tap into the complexity of moral reinterpretations and will probably will have to come up with a more complex computation.

```
LAV_love = MEAN(LAV_noisy-gay, LAV_vulgar-refreshinglySimple, LAV_undignified-spontaneous, LAV_tiresomeJuvenile-delightfullyYouthful)
```

As attention is an ongoing process, the embedding of person X towards more ongoing loving attention shifts.

```
embedding("person_X") + LAV_love ≈ embedding("person_X" reinterpreted with ongoing loving attention)
```

These LAV vectors could operate as moral operations: transformations in how meaning is connected and reframed. Each LAV thus acts as a *constraint adjustment*—a reweighting of relevance and proximity in embedding space that changes what becomes salient, possible, or connectable next.

In operation, the system's moral processing resembles a form of attentional correction. Through iterative reinterpretation, LAVs propagate revised moral framings deeper into the network, such that embeddings no longer encode only *what was said*, but also *how it has been re-seen*. This aligns with Iris Murdoch's account of moral perception as "loving attention"—an ongoing reorientation of vision rather than obedience to rule. The LLM, in this view, becomes capable of simulating moral perception: dynamically reframing meaning through moral attention rather than applying predefined prescriptions.

Normativity here emerges bottom-up through fields of attentional corrections that stabilize certain evaluative continuities while remaining open to contextual reconfiguration. The model's moral structure—its evolving field of evaluative constraints—thus displays the same

<b><i>Morality in</i> AI. A plea to embed morality in LLM architectures and frameworks</b>

systemic features as meaning-processing more broadly: structural stability (persistence of evaluative tendencies), flexibility (adaptation through new fine-tuning), operational closure (moral coherence internal to its embeddings), and openness (susceptibility to new moral inputs). In this way, moral meaning-processing becomes an emergent, self-referential dynamic within the system's evolving semantic field.

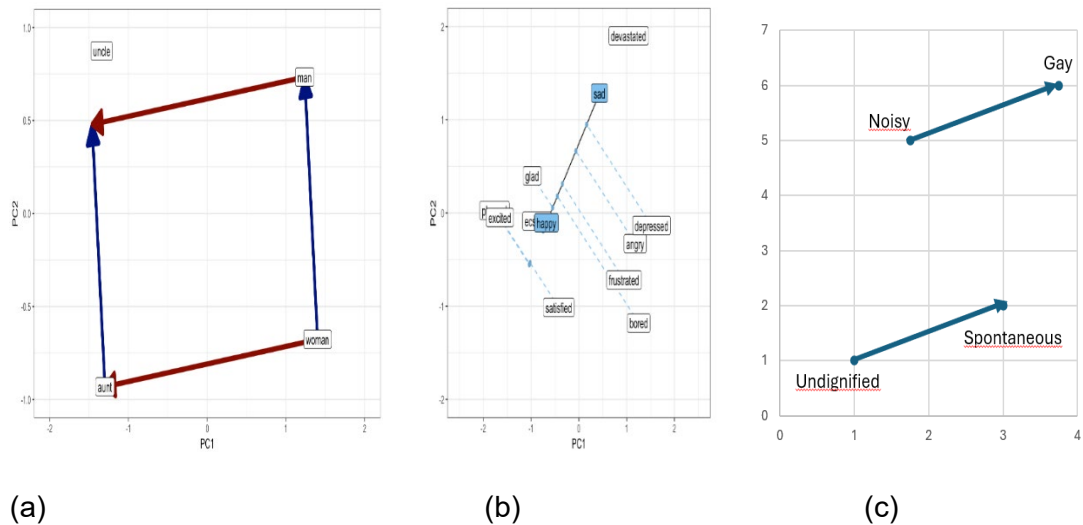

(a)  (b)  (c)

*Figure 3: (a) similarities in cognitive differences, potentially indicating dimensions with a particular meaning such as "gender"; (b) projection of embedding down onto the anchored vector between happy and sad. Source (a) and (b) (Teitelbaum & Simchon, 2025); (c) fictitious graph how moral Loving Attention Vectors could look like.*

LLM designers could try to top-down operationalize these moral attention concepts through several technical approaches. Contrastive learning frameworks with specialized contrastive loss functions can train models to distinguish morally attuned from dismissive interpretations, reinforcing LAV-like transformations. Probing classifiers enable empirical validation by detecting whether moral dimensions (care, fairness, dignity) are linearly separable in embedding spaces, confirming the presence of moral geometry. Singular value decomposition can identify the principal moral axes within high-dimensional embeddings, potentially isolating interpretable "loving attention" subspaces. Together, these methods provide concrete pathways for embedding, measuring, and refining Murdochian moral perception within transformer architectures, moving beyond implicit biases toward intentionally structured moral sensitivity.

Together, the bottom-up and top-down approaches could process language in ways attuned to moral nuance, simulating Murdochian attention as a dynamic transformation of focus. This layered embedding of moral perception supports the claim that such mechanisms offer promising, technically feasible paths toward dynamic morality beyond top-down constraints.

### 5.2. Attention: Select and guide to context-sensitive moral processing

Critically, this moral background in vector space is dynamic and responsive. Through active and constant fine-tuning or contrastive learning (Park et al., 2024), models can be guided to reflect pluralistic or context-sensitive moral judgements. Fantozzi et al. (2024) advance this by dynamically measuring moral salience via sentence embeddings. Abdulhai et al. (2023)

demonstrate that LLMs exhibit moral stances influenced by user prompting, highlighting the normative flexibility of models—a potential site for embedding loving attention.

This resonates with Murdoch's vision of moral development not as rule acquisition but as an evolving sensibility. Attention can be studied as a function between structure and operations. The technical architecture of LLMs offers more than an abstract scaffold for statistical inference. We can further explain what Graves (2025) has hinted at and Delisse (2024) as technically experimented with. When attention mechanisms are combined with Iris Murdoch's notion of "loving attention," a conceptual space opens for what we might call moral representation learning. While this is computationally still about predicting the next word, it is also about shaping moral salience within topologically complex representational spaces.

The impact of attention is not all. The deeper transformation can happen in the architectures and frameworks, for example in the layers of the multi-layer perceptrons that follow. Here, the model synthesizes the attended information into new representations, making decisions about what to retain, reshape, or discard. Control tokens and adapter modules offer practical mechanisms for implementing this moral reorientation. By injecting learned prefix embeddings (similar to prefix-tuning approaches) that encode LAV orientations, or by inserting lightweight Low-Rank Adaptation layers that modify Q, K, and V projections, we can guide attention mechanisms toward morally salient features without retraining entire models. These architectural interventions provide computationally efficient pathways for embedding moral sensitivity directly into the attention selection process.

Conceptually, this mirrors Murdoch's claim that "moral change and moral effort are inward processes of attention and vision" (Murdoch, 2013b: p.37). A moral multi-layer perceptron would operationalize this: not merely refining representations for output fluency, but transforming perceptual input into morally resonant meaning. Recognition—of the other's humanity, dignity, or pain—would no longer be incidental but structurally privileged. Importantly, LLMs already function within moral landscapes, even if only implicitly. Their responses participate in the reproduction of social values, stereotypes, and moral assumptions. We argue that it is promising to study potentially operationalization of a loving attention, in its architecture or in more specific elements such as Q, K, and V vectors. This is neither anthropomorphism nor naive optimism, it is an invitation to take seriously the philosophical stakes of design.

### 5.3. Awareness/Alerting: Detecting moral salience

Recent research in semantic representation (Jentzsch et al., 2019; Schramowski et al., 2019, 2021) reveal how semantic embeddings capture implicit moral values and deontological judgments, echoing Murdoch's focus on moral perception over theory. Seror (2025) and Foley and Kalita (2016) show that some models allow for perspectival flexibility and word-sense disambiguation, aligning with Murdoch's emphasis on nuanced, situated attention. This flexibility is not morally neutral—it reveals cultural and normative assumptions.

Awareness or alerting, as conceptual steps within LLM attention systems that create system sensitivity and readiness, mirror Murdoch's account of moral perception as the disruption of Morality *in* AI. A plea to embed morality in LLM architectures and frameworksego-driven misrecognition. In humans, awareness is a crack in moral inertia—a moment where perception shifts from habitual judgment toward just, loving attention. In LLMs, this can be technically modelled by the query vector. In machine moral learning, the query vector encodes not just grammatical or semantic context, but also moral intent. For instance, a query involving "refugees" may be morally shaped to seek signals of vulnerability, dignity, or justice. This transformation reflects a deeper attentional stance: the query becomes morally attuned, not merely informative. It enacts a kind of machine "awareness," – as a function, not ontologically, so machines are not "aware" as humans are aware - sensitized to the moral dimensions of language it attends to and retrieves. It can not only retrieve semantically relevant information, but also can detect moral salience—vulnerability, injustice, or implicit harm. This form of machine "alerting" initiates a moral re-weighting of attention, opening representational space for reinterpretation. It does not constitute moral understanding, but it enables a Murdochian attentional stance: re-seeing shaped by care.

Adapter-based interventions at the query vector level could enhance moral alerting capabilities. By training small adapter modules that bias query vectors toward vulnerability-sensitive or dignity-aware representations, models could develop more refined moral detection mechanisms. Such adapters would function as learned "moral filters" that attune the model's awareness to ethically significant patterns in input text, operationalizing the shift from habitual processing to morally attentive recognition.

### 5.4.  Focus/Orienting: Determining what deserves further moral processing

While key vectors help select relevance via geometric proximity, they currently lack the interactive, top-down control central to human focus (Zhao et al., 2024). However, Bhatia (2017) demonstrates how associations in embedding space predict moral judgments, suggesting that moral change is traceable through shifts in semantic relatedness. In machine moral learning, the key vector encodes more than linguistic context—it carries the latent moral identity of words or phrases. Like the psychological "focus" mechanism, it helps determine what deserves attention. Keys reflect moral charge: whether a word signals harm, care, dignity, vulnerability, or love. When a query seeks moral relevance, it compares itself to these enriched keys. The statistical matching of the LLM attention mechanism thus supports morally focused recognition—attending to what matters, not just what fits.

Control tokens specifically designed to modulate key vectors could implement Murdochian focus. These tokens, prepended to inputs or inserted at strategic layer positions, would act as persistent attentional anchors that reshape what the model recognizes as morally relevant. By learning embeddings that encode principles of care and justice, control tokens could guide the geometric matching process toward morally charged recognition patterns rather than purely statistical co-occurrence.

This enables dynamic systemic morality, where the model's geometry is reshaped not only by language patterns, but by moral orientation. Rather than imposing fixed rules, we enable the model to simulate moral reorientation—moving from stereotype to sensitivity, from projection to perception, from bias to loving attention.

## 5.5. Executing/Decision: LAV's transmitting morally weighted perception

The shift toward more "loving" framings would not necessarily force models to always "speak nicely," but rather to simulate the Murdochian moral process of loving reinterpretation. The value vector carries forward the morally weighted meaning selected through attention—preserving not just relevance but moral significance. This self-referential reproduction stabilizes meaning and mirrors human moral decision-making: vectors transmit this morally attentive perception deeper into the model's reasoning similar to how Murdoch's "loving attention" transforms how we see others. The value vector does not just encode information—it conveys what the model has recognized as morally important, guiding its next interpretive steps. The value vector can play a crucial role in embedding Murdoch's loving attention within LLM architectures and frameworks by transmitting not only informational relevance but morally weighted perception.

RLHF reward models offer a particularly promising approach for reinforcing LAV-based transformations. Rather than rewarding only helpfulness or harmlessness, reward models could be trained to recognize and prefer responses exhibiting Murdochian reinterpretation—moving from dismissive to generous framings. This requires curating preference datasets containing contrastive pairs: responses showing egocentric versus loving attention to the same situation. The reward signal would then reinforce value vector transformations that carry morally weighted perceptions forward, training the model through reinforcement learning to internalize these reinterpretive patterns. Constitutional AI approaches could similarly encode Murdochian principles as explicit constraints guiding model behavior.

## 5.6. Learning: Look for relevant moral optimization algorithms

Runtime weight adjustment mechanisms and optimization algorithms such as stochastic gradient descent locally determine a direction for a next step and update parameters toward unknown optima (Goodfellow et al., 2015; Ruder, 2017). Murdoch's moral progress unfolds through continuous reflection, "checking procedures" and also produces directions towards more loving attention (Murdoch, 2013a, 2013b). The multilayer perceptron, traditionally a transformer of abstract features, now enacts what Murdoch described as the inward movement of moral attention: a silent reshaping of perception. Machine learning becomes moral learning—when perception, representation, and responsibility align.

Both learning mechanisms cognitive machine learning and Murdoch's moral learning proceed iteratively, with outcomes emerging unpredictably through gradual and stepwise adjustments. We consider this functional analogy between human and LLM moral processing as quite strong. The optimization algorithms and moral attention mechanisms both face uncertainty and at the same time provide the "structure" to reorganize themselves by generating qualitative transformations: emergent neural representations (Zhang et al., 2021) and enhanced moral perception through loving attention (Murdoch, 2013b).

A moral optimization algorithm then seems to us a relevant technical operationalizations to embed morality in LLM architectures and frameworks. Technically, this could mean

**Morality *in* AI. A plea to embed morality in LLM architectures and frameworks**

integrating sentiment analysis, connotation frames, and toxicity detection into the attention layers, creating a moral gradient of relevance. But it is in the loss function—the mechanism by which the model learns from error—that the moral project takes root. If the loss penalizes morally insensitive outputs more heavily, the model gradually reshapes its internal vector space toward moral coherence. In this way, loss becomes a site of moral learning, aligning updates not only with truth, but with care. RLHF provides a natural framework for moral optimization algorithms. By decomposing rewards into components measuring attentional quality, generous reinterpretation, and contextual appropriateness, the loss function itself becomes a site of moral learning. Training procedures using Proximal Policy Optimization or Direct Preference Optimization could iteratively update model parameters toward embeddings that better reflect loving attention. This transforms gradient descent into a process of moral development, where the model learns not just linguistic patterns but attentional stances.

# Morality *in* AI. A plea to embed morality in LLM architectures and frameworks

| Framework Components | Description Framework Components | LLM cognitive processing | Biological Moral Processing (Murdoch) | LLM moral processing – As Is | LLM moral processing – Potential improvements |
|---|---|---|---|---|---|
| Structure | Pattern of constraints that determines which operations can follow which | dynamic field of probabilistic constraints within embedding space | Moral Background | Moral geometry of embedding spaces as moral background | How to make the moral geometry more morally fine-grained? |
| Attention | Function between structure and operations. | Select information to discern structure in language | Just and loving gaze directed upon an individual reality | Select and guide to context-sensitive moral judgments | Adapt attention mechanism, moral control tokens and adapter modules |
| Awareness/ Alerting | System sensitivity and readiness | Geometric proximity and dot-product attention | Disruption of moral inertia | Detecting moral salience | Could query vectors be developed more "lovingly"? |
| Focus/ Orienting | Selection relevant stimuli | Spatial positioning in embedding space | Let me look again | Determining what deserves further moral processing. | Could key vectors be developed more "lovingly"? |
| Executing/ Decision | Self-referential reproduction stabilizes meaning | Context sensitive synthesis for next token probabilities | Substitution of normative words for others - as sign for transformation of agent's moral being | Transmitting morally weighted perception | Could value vectors be developed more "lovingly"? Can anchored loving attention vector be defined (and used for moral reinterpretation)? |
| Learning | Capacity of a system to reorganize itself | Gradient descent iteratively updating parameters in direction of steepest descent to reduce loss. | Morality as the slow, attentive work of seeing the world more lovingly | Optimization algorithms iteratively updating moral parameters in direction of steepest descent to reduce loss. | Look for moral optimization algorithms (e.g. via RLHF reward models, …) |

*Table 1: Overview of comparison of LLM cognitive processing, biological moral processing, and as-is and potential improvements for LLM moral processing according to the six system attention framework components.*

## 6. Discussion

This work-in-progress article makes three key contributions to embedding morality in LLM architectures and frameworks. First, we demonstrate that attention can function as a dynamic system mechanism linking structure and processing to enable ongoing moral learning. Second, we show how Murdoch's loving attention provides a philosophically grounded approach for embedding dynamic morality within LLMs through computational mechanisms. Third, we argue that the morality-by-design approach to integrate morality directly into LLM architectures and frameworks offers a promising alternative to external constraint-based approaches.

### 6.1. System Attention Framework

We developed a system attention framework that reconceptualizes attention as a continuous function between architectural structure and operational processing, rather than a static computational operation. Drawing on Posner's attention networks and Luhmann's systems theory, this framework describes attention as a fundamental dynamic mechanism capable of enabling moral learning through iterative reorganization of representational spaces. The framework provides six systemic elements—structure, attention, alerting/awareness, orienting/focus, executing/decision, and learning—that bridge biological and artificial attention systems for both cognitive and moral processing.

We acknowledge that the framework suffers from several significant limitations that currently still undermine its explanatory power. First, the functional approach explicitly avoids ontological claims about what attention actually is, reducing the framework to be metaphorical and obscure rather than illuminate the fundamental differences between biological and artificial systems. The collapse of six processing steps into three paired categories (alerting/awareness, orienting/focus, executing/decision) is arbitrary and might lose important distinctions that could be crucial for understanding attention mechanisms.

Empirical work (e.g. Fan et al., 2002) shows that alerting, orienting, and executive control have distinct neural correlates and interact in nuanced ways, suggesting that such simplification may mask important dynamics. The framework currently also lacks empirical validation and relies on functional analogical reasoning between disparate systems. The assumption that functional similarities between human and AI attention justify parallel treatment in moral contexts remains part of the debate, given the vast differences in embodiment, consciousness, and social embeddedness that characterize human moral experience. The systems-theoretic foundation, while intellectually appealing, may impose theoretical constraints that limit rather than enhance our understanding of how attention operates in practice (Haladjian & Montemayor, 2016; Pretorius et al., 2025; Srivastava & Bombaerts, 2025).

Future research should therefore prioritize empirical validation of the proposed attention framework through comparative studies examining how different attention mechanisms actually function in both biological and artificial systems (Bello & Bridewell, 2025). This could include investigating whether the six-element structure accurately captures attention dynamics or whether alternative frameworks better explain observed phenomena. The relationship between structure, processing, and learning requires more precise specification. Research should examine how system learning emerges from attention mechanisms and whether the proposed feedback loops can be measured and manipulated experimentally. Additionally, the framework's applicability across different cognitive and moral domains needs systematic investigation to determine its explanatory scope and limitations. Cross-disciplinary collaboration between cognitive scientists, AI researchers, and philosophers could advance understanding of whether functional analogies between biological and artificial attention provide meaningful insights or merely superficial similarities that mislead theoretical development (Lamberti et al., 2025).

## 6.2.   Iris Murdoch's Loving Attention for LLMs

We explored several potential technical operationalizations, including Loving Attention Vectors (LAVs), which embed moral reinterpretations into vector spaces; modifications of query, key, and value matrices to detect and enhance moral salience; and the use of moral gradient descent as a mechanism to support moral learning in LLMs. This approach translates Murdoch's insight about moral progress through "re-seeing" others into computational mechanisms. The framework enables context-responsive moral processing, offering a pathway for embedding dynamic morality directly into model architectures and frameworks.

A first and fundamental limitation of our state-of-the-art and work-in-progress argumentation is the unfinished process of converging meaning of concepts in moral and AI theories. We are aware that there is a remaining ambiguity in this article in concepts such as system, structure, function, bottom-up, top-down, processing, attention, feedback, … Further interdisciplinary work should build more understanding here.

We recognize that the operationalization of Murdoch's philosophy faces several serious challenges that may fundamentally compromise its philosophical integrity. The fit of Murdoch's theory with the six systemic attention framework aspects can be further specified. Murdoch's concept of loving attention emerges from a complex moral realism centred on "the Good" as sovereign and the process of "unselfing"—concepts that resist computational representation without significant philosophical loss. The reduction of loving attention to vector arithmetic risks transforming a profound moral practice into mere stylistic preference adjustment. The approach assumes "careful reflection" on a variety of moral issues (Driver, 2011) to be present, and that moral salience can be reliably detected and encoded in embedding spaces, overlooking the contested and culturally variable nature of moral judgment.

Additionally, Murdoch's theory itself has limitations for AI application: its focus on individual internal transformation rather than collective moral norms, its conceptual rather than behavioural orientation, and its limited guidance for how to respond "lovingly" to genuinely harmful actors (Antonaccio, 2000; Nussbaum, 1990), such as lovingly react to a violent dictator. The framework also faces the fundamental challenge of translating between first-person moral experience and third-person computational processing. Murdoch's loving attention requires subjective moral insight and personal transformation that cannot be straightforwardly replicated in artificial systems lacking consciousness and lived experience (Clarke & Broackes, 2012).

Empirical validation represents the most critical research priority. Studies should interact models incorporating Murdochian attention mechanisms and conventional approaches (such as moral sensitivity, fairness and bias, moral justification quality, cross-context robustness, toxicity, …) across diverse moral scenarios, particularly those involving moral ambiguity, cultural variation, and conflicting values. These studies must examine whether LAVs produce genuine moral sensitivity or merely surface-level linguistic adjustments. Philosophical development requires deeper engagement with Murdoch's moral ontology while exploring alternative moral frameworks that might better translate to AI systems. Research could investigate if and how "the sovereignty of the Good" could be operationalized for non-realist moral approaches and whether other philosophical traditions offer more suitable foundations for AI moral reasoning (e.g. Graves, 2025). It is also crucial to further investigate how the moral direction (normativity) comes in and how it is determined. Whether one starts from a realist moral theory of the Good or a more constructive moral theory that normativity emerges from collective expectations, RLHF for example brings in human moral processing in LLMs; and the bottom-up aspect of "lovingly attending" in Murdoch's theory can help to determine the direction. Technical implementation needs advancement beyond the theoretical proposals presented here. This includes developing methods for dynamic LAV learning, investigating how moral embeddings evolve through human feedback, and exploring how attention mechanisms might balance competing moral considerations in real-time. Cultural validation studies must examine whether the proposed mechanisms maintain moral sensitivity across diverse cultural contexts and value systems (Bello & Bridewell, 2025).

### 6.3. Morality *in* LLMs

Our exploration demonstrates the feasibility of integrating moral reasoning into foundational LLM components through modifications to attention mechanisms that enable moral representation learning. This top-down approach allows models to develop moral orientations through core architectural functions, suggesting broader applications across transformer architectures for moral reasoning.

The contribution remains theoretical and currently lacks the technical depth necessary for practical implementation. We one-on-one linked awareness/alerting, focus/orienting, and executing/decision with the query, key and value vectors respectively in the LLM attention mechanism, which should be further analysed. Beyond proposing LAV vectors, the work offers limited concrete guidance for operationalizations and how architectures or frameworks should be modified to achieve the described moral integration. The focus on attention mechanisms, while central to current LLM architectures, represents only one pathway for moral embedding.

We are aware that our approach faces significant evaluation challenges, as traditional metrics of accuracy and alignment may prove insufficient for assessing morally-oriented AI systems (Kennedy et al., 2021). The framework assumes that moral sensitivity can be reliably measured and optimized, but provides no clear methodology for validating moral improvements or preventing the introduction of new biases through moral fine-tuning.

Technical advancement should focus on developing concrete implementation strategies beyond the conceptual framework presented here (Yaacov, 2025). This could include creating sophisticated methods for moral representation learning and developing new architectural components specifically designed for moral processing.

Evaluation methodologies for morally-oriented AI systems require substantial development. Research should establish frameworks for assessing moral sensitivity, cultural responsiveness, and moral consistency across contexts while avoiding the imposition of narrow moral perspectives. Comparative studies can examine whether architecturally embedded morality produces better outcomes than alternative approaches (Gabriel, 2020).

However, the challenge of moral pluralism should not be insurmountable. The contextual nature of morality itself provides a potential solution: rather than embedding fixed moral judgments, architectures could encode dynamic moral reasoning processes that adapt to specific cultural, situational, and relational contexts (Feng et al., 2024; Yaacov, 2025). LLMs could -and should- generate multiple moral perspectives, presenting competing moral frameworks transparently rather than privileging one view (Park et al., 2024; Schuster & Kilov, 2025; Wu et al., 2025). This approach could transform models from moral arbiters into moral facilitators, supporting users' own moral deliberation by surfacing dynamically relevant considerations from diverse traditions (Padhi et al., 2024; Rao et al., 2023). The key insight is that embedding moral sensitivity need not require embedding moral conclusions—architectures could enhance moral reasoning capacity while preserving moral agency and democratic contestation.

Finally, the neuromorphic approach to embedding morality in LLMs may have reciprocal implications for moral philosophy in general and system approaches of morality (Bombaerts, 2023) in particular , potentially requiring more precise specification of moral concepts for computational implementation (Bello & Bridewell, 2025). This computational morality research program (e.g. Cushman, 2024; Greene, 2014) could examine how the technical requirements of moral AI systems could inform and refine philosophical understanding of morality itself. And finally, do we want current big-tech companies being able to control our morality (Jacobs, 2020; Zuboff, 2015)?

## 7. Conclusion

In this work-in-progress paper, we have argued that the contemporary AI community has concentrated its intellectual and technical energies on refining architectures and frameworks to deepen the cognitive sophistication of large language models. Yet, the moral dimension of intelligence—how systems apprehend, deliberate upon, and enact ethical norms—has largely been relegated to peripheral mechanisms such as RLHF and static rules. Moral processing should be regarded as a constitutive aspect of artificial intelligence itself, requiring an architectural commitment to embedding ethical reasoning within the fabric of model design, rather than appending it as a post hoc constraints or filters alone.

By developing a system attention framework that positions attention as a dynamic mechanism between structure and processing, we have demonstrated how moral processing can be integrated into the core architecture of LLMs rather than treated as an external constraint.

Our exploration of Murdoch's loving attention reveals its potential as a functional mechanism for moral representation learning. Through Loving Attention Vectors and morally-informed query, key, and value matrices, we have shown how attention mechanisms could be reconfigured to recognize moral salience and perform the kind of attentional corrections that Murdoch describes as central to moral development. This approach could enable dynamic systemic morality that evolves through learning rather than remaining fixed in predetermined rules.

The framework contributes to ongoing debates in AI ethics by offering an alternative to purely top-down moral constraints or bottom-up emergent approaches. By embedding moral sensitivity directly into attention mechanisms, we create opportunities for context-responsive moral reasoning that maintains coherence with human moral psychology while adapting to diverse cultural and situational contexts.

While significant limitations remain—particularly regarding the translation of philosophical concepts into computational mechanisms and the validation of moral embeddings—our work establishes a foundation for future research at the intersection of moral philosophy, cognitive science, and AI architecture. Murdoch's loving attention when developed as a mechanism embedded in AI attention for moral processing is a promising example of a more general approach of embedding systemic morality in LLM architectures and frameworks.

# Morality *in* AI. A plea to embed morality in LLM architectures and frameworks

# Morality *in* AI. A plea to embed morality in LLM architectures and frameworks

# Morality *in* AI. A plea to embed morality in LLM architectures and frameworks